\documentclass[pre,superscriptaddress,amsmath,amssymb,nofootinbib,floatfix,onecolumn,notitlepage]{revtex4-1}
\usepackage{amsfonts, amsmath, amssymb, amsthm, array, eucal, mathtools, xfrac}
\usepackage{booktabs, array, color}
\usepackage{natbib}
\newlength\fwidth
\newlength\fheight
\usepackage{graphicx}
\usepackage{siunitx}

\begin{document}
	
	\title{The autophoretic torus}
	
	\author{Lasse C. Schmieding}
		\affiliation{Department of Applied Mathematics and Theoretical Physics,\\ University of Cambridge,
	Wilberforce Road, Cambridge CB3 0WA, UK.}
	\author{Eric Lauga}
		\affiliation{Department of Applied Mathematics and Theoretical Physics,\\ University of Cambridge,
	Wilberforce Road, Cambridge CB3 0WA, UK.}
	\author{Thomas D. Montenegro-Johnson}
	\email{t.d.johnson@bham.ac.uk}
	\affiliation{Department of Applied Mathematics and Theoretical Physics,\\ University of Cambridge,
	Wilberforce Road, Cambridge CB3 0WA, UK.}
	\affiliation{School of Mathematics, University of Birmingham, \\
Edgbaston, Birmingham, B15 2TT, UK.}
	
	\date{\today}
	
	\begin{abstract}

{Phoretic swimmers provide} new avenues to study non-equilibrium
statistical physics and are also hailed as a promising technology for
bioengineering at the cellular scale. Exact solutions for the locomotion of such
swimmers have been restricted so far to spheroidal shapes. In this paper we
solve for the flow induced by the canonical non-simply connected shape, namely
an axisymmetric phoretic torus. The analytical solution takes the form of an
infinite series solution, which we validate against boundary element
computations.  For a {torus} of uniform chemical activity, confinement effects in the
hole allow the torus to act as a pump, which we optimize subject to
fixed particle surface area. Under the same constraint, we next characterize the
fastest swimming Janus torus for a variety of assumptions on the surface
chemistry. Perhaps surprisingly, none of the optimal tori occur in the limit
where the central hole vanishes.
				
	\end{abstract}
	
	\maketitle
	
	\section{\label{sec:1}Introduction}
	
{Autophoretic microswimmers are artificial microscale particles that self propel via 
slip flows at their surface created through self-generated, rather than
externally imposed~\cite{anderson1991,khair2013}, field gradients such as   heat~\cite{jiang2010,bickel2013} or solute
concentration~\cite{golestanian2005}. Such particles have potential
biomedical~\cite{BJN1} and microfluidics applications~\citep{maggi2016self}, and
may perform intricate microscale tasks, for example directed cargo transport and
assembly~\citep{popescu2011,baraban2012transport}.}

{In this study, we  focus on neutral solute
self-diffusiophoresis, in the absence of electrokinetic
effects~\cite{ebbens2014,brown2014}, whereby} solute chemical reaction
differentially catalysed at the swimmer surface leads to local concentration
gradients around the swimmer; this property is know as the swimmer's
\emph{activity}. Local pressure imbalances arising from this differential
interaction between the particle surface and the solute in a thin layer drive
surface slip flows, propelling the swimmer forward; this property is known as
the swimmer's \emph{mobility} \cite{anderson1991}.

{The trajectory of autophoretic microswimmers thus results from the coupled
interactions of solute concentration and hydrodynamics, which are in turn
strongly influenced by physical conditions such as particle
shape~\cite{popescu2010} and domain boundaries. As such, autophoretic
microswimmers exhibit a range of remarkable and complex dynamics, such as
self-assembly with neighbouring particles~\cite{sharifimood2016} into ``living
crystals''~\cite{palacci2013} or rotors and
swimmers~\cite{wykes2016dynamic}, phase separation~\cite{IB1}, swarming
behaviour~\cite{thutupalli2011}, boundary-following~\cite{das2015boundaries,simmchen2016topographical} and  
rheotaxis~\cite{uspal2015rheotaxis}.}

{The self-generation of concentration gradients is typically
achieved either through chemical patterning~\cite{paxton2004}, solute
advection~\cite{michelin2013spontaneous}, or geometric effects such as varying
particle curvature~\cite{shklyaev2014non} or through confinement interactions,
for instance with a boundary wall~\cite{uspal2015self,YI1,mozaffari2016self}.}
The canonical patterned autophoretic microswimmer is the Janus
particle~\cite{AW1}; an inert sphere or rod, for instance a polymer, is
half-coated in a catalyst for the solute, such as platinum in hydrogen
peroxide~\cite{howse2007,Ebbens2011}.

Previous theoretical studies on autophoretic motion in three dimensions (3D) have focused on simply
connected particle geometries, such as spheres~\cite{SM1} or rods~\cite{RG1}.
{Theoretical and numerical studies examining the effects of
confinement have typically focused upon the prototypical problem of  a Janus particle over a plane
boundary~\cite{uspal2015self,YI1,mozaffari2016self}}, or in the form of
multi-particle interactions~\cite{thakur2012collective}.  In this paper, we
explore the phoretic motion of the canonical shape which is not simply
connected: the torus. The central hole of the torus provides an example of
{intrinsic} geometric confinement~\cite{SM2}, allowing the
consequences of this physical effect to be explored, {and
optimized~\cite{popescu2011}}, within an analytical framework. Confinement leads
to locally higher concentrations, generating a pumping flow even for uniform
surface chemistry. {This is similar to the method of generating
concentration gradients via changes in surface curvature leading to swimming for axisymmetric
shapes where front-back symmetry is broken~\cite{shklyaev2014non}.} This is in
contrast to isolated spherical particles, which can only pump flow via chemical
patterning (eg Janus particles~\cite{golestanian2005,RG1}), interactions with
neighbours~\cite{SM1}, or solute advection~\cite{michelin2013spontaneous}.

{In this paper, we consider  the continuum framework for phoretic motion at zero
P\'eclet number developed by Golestanian et al~\cite{golestanian2005}; for an
alternative framework, see~\cite{cordova2008,brady2011,cordova2013}. Using}
toroidal coordinates, we derive the analytical solution for the phoretic torus
with axisymmetric boundary conditions in terms of an infinite sum of Legendre
polynomials of half-integer order. We consider several cases, with specific
solutions validated against 3D regularized boundary element code~\cite{TDMJ1}.
We find optimal tori of fixed surface area, an important constraint for
controlling reaction rates~\cite{TDMJ2}, that either produce the most pumping or
maximize swimming speed. In contrast to previous studies~\cite{SM1,SM2,ML1},
these optima do not occur when confinement is maximized (i.e.~where the hole is
infinitely small). 
	
\section{\label{sec:2}The autophoretic Torus}
	
\subsection{\label{subsec:2.1}Continuum framework}
		
We consider the autophoretic motion of a torus $\mathcal{S}$, with axisymmetric
surface chemistry, in the continuum framework~\cite{golestanian2005,RG1}. The torus
interacts with a solute fuel $S$. Catalysis at the surface of the torus converts
the fuel to a product $P$, which has a local concentration $c$ and is dissolved
in a fluid of dynamic viscosity ${\mu}$ and a density $\rho$. The torus has
surface activity $\mathcal{A}(\mathbf{x})$, such that the flux of product
through the surface is
\begin{equation}
\label{eq:1} 
-D \mathbf{n}\cdot\bm{\nabla}c = A(\mathbf{x}),
			\end{equation}
			where $\mathbf{n}$ is the unit normal pointing into the fluid, $D$ denotes the diffusivity of the product, and $c$ is the concentration difference against the far field. If the P\'eclet number $\text{Pe} = \mathcal{U}\mathcal{R}/D$ is small, with $\mathcal{U}$ and $\mathcal{R}$ being characteristic velocity and length scales of the problem, the concentration $c$ satisfies
\begin{equation}
\label{eq:2} \nabla^2 c = 0 \quad \text{ outside $\mathcal{S}$},\quad c \to 0 \quad \text{as $r \to \infty$},
\end{equation}
			together with the Neumann boundary
condition~(\ref{eq:1}), {appropriate when the ratio of diffusive to reactive
timescales, the Damk\"{o}hler number, is small~\cite{michelin2014phoretic}}. The interaction between $\mathcal{S}$ and the solute can be modelled as giving rise to a slip velocity along the surface
			\begin{equation}
				\label{eq:4} \mathbf{u}_s = M(\mathbf{x})\left( \mathbf{1} - \mathbf{n}\mathbf{n} \right) \cdot \bm{\nabla}c \quad \text{on $\mathcal{S}$},
			\end{equation}
			with $\mathcal{M}(\mathbf{x})$ the local surface mobility of $\mathcal{S}$. Neglecting inertial effects, due to typical particle size and flow speeds being small, the flow around the torus is then governed by the Stokes flow equations
\begin{subequations}
			\begin{align}
				\label{eq:5}  {\mu} \nabla^2 \mathbf{u} &= \bm{\nabla}p, \\	
				\label{eq:6}  \bm{\nabla} \cdot \mathbf{u} &= 0.
			\end{align}
\end{subequations}
			For the axisymmmetric torus, the swimming velocity is
$\mathbf{U} = U \mathbf{e}_{z}$, which can be obtained by requiring there to be
zero net force on $\mathcal{S}$. Working in the reference frame centered on the
torus, the boundary conditions for the Stokes flow problem become
\begin{equation}
\label{eq:7} \mathbf{u} = \mathbf{u}_s \text{ on $\mathcal{S}$,}\quad
\mathbf{u} \sim -U\mathbf{e}_z \text{ as $r \to \infty$.}
\end{equation}
{Taking $\CMcal{A}$ and $\CMcal{M}$ to be typical magnitudes of
the surface activity and mobility respectively, we nondimensionalize the problem
by setting $\CMcal{A} \mathcal{R} / D$ to be the characteristic size of
concentration fluctuations, $\CMcal{M}\CMcal{A} / D$ the characteristic size of
the slip velocity, and ${\mu} \CMcal{M} \CMcal{A} / (\mathcal{R} D)$
as the characteristic size of the dynamic pressure.}
		
		\subsection{\label{subsec:2.2}Toroidal coordinates}
		
\begin{figure}[tb]
\begin{center}
\includegraphics{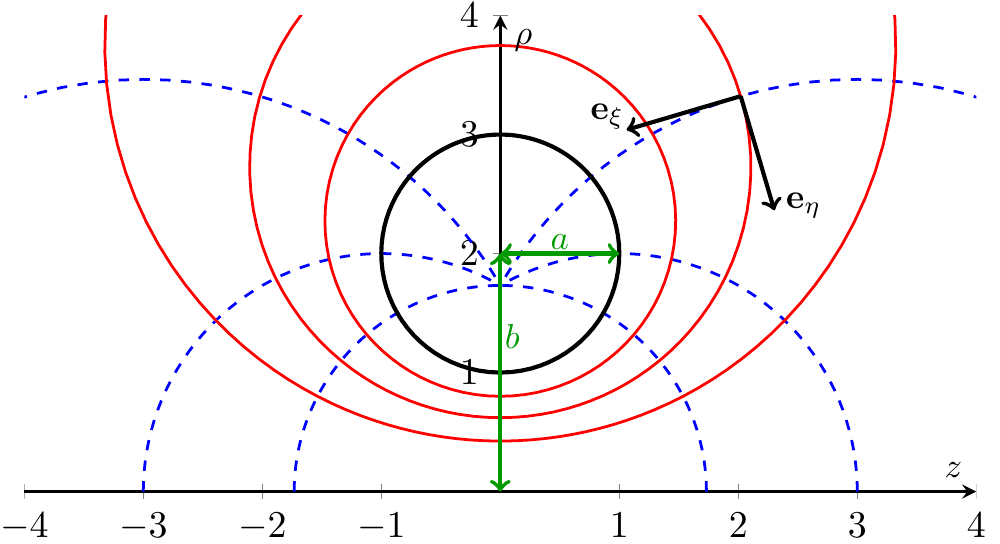}
\end{center}
\vspace{-0.3cm}
\caption{Toroidal coordinates (redrawn from Ref.~\cite{AML1}) for a torus with $a = 1, b = 2$, with lines of
constant $\xi$ (red, solid) and lines of constant $\eta$ (blue, dashed). The
contour $\xi = \xi_0$ represents the boundary of the torus and is shown by the
thick black circle. The unit vectors $\mathbf{e}_{\xi},\mathbf{e}_{\eta}$ are shown at $\xi =
0.75,\eta=\pi/6$.}
\label{fig:coords}
\end{figure}

			Let $(\rho, \phi, z)$ be standard cylindrical coordinates. For this problem we introduce  toroidal coordinates~\cite{AML1, PM1} $(\xi, \eta, \phi)$, which are related to cylindrical coordinates through the transformation
\begin{equation}
\label{eq:9} \rho = d \frac{\sinh{\xi}}{\cosh{\xi}-\cos{\eta}},\quad z = d \frac{\sin{\eta}}{\cosh{\xi} - \cos{\eta}}, 
\end{equation}
			where $d > 0$, $0 \leq \xi < \infty$ and $0 \leq \eta <2
\pi$. In this coordinate system, shown in Fig.~\ref{fig:coords}, curves of constant $\xi$ correspond to circles
in the $(\rho,z)$ plane with radius $a = d/\sinh{\xi}$, centred at $(b,0)$,
where $b = d \cosh{\xi}/\sinh{\xi}$. Rotating by $2 \pi$ around the $z$ axis,
the curves of constant $\xi$ become tori. The scale factors $h_{i}$ and the unit
vectors $\mathbf{e}_i$ are given by
			\begin{equation}
				\label{eq:11} 
h_{i}\mathbf{e}_{i} = \frac{\partial \mathbf{x}}{\partial q_{i}},\quad\mbox{so
that}\quad h_{\xi} = h_{\eta} = \frac{d}{\cosh{\xi}-\cos{\eta}},
{\ h_\phi = \frac{d\sinh\xi}{\cosh\xi - \cos\eta},}
			\end{equation}
			and the unit vectors $\mathbf{e}_{\xi}$ and $\mathbf{e}_{\eta}$ are
\begin{subequations}
			\begin{align}
				\label{eq:12}  \mathbf{e}_{\xi} &= \frac{1-\cosh{\xi}\cos{\eta}}{\cosh{\xi}-\cos{\eta}}\mathbf{e}_{\rho} - \frac{\sin{\eta}\sinh{\xi}}{\cosh{\xi}-\cos{\eta}}\mathbf{e}_z, \\
				\label{eq:13}  \mathbf{e}_{\eta} &= -\frac{\sin{\eta}\sinh{\xi}}{\cosh{\xi}-\cos{\eta}}\mathbf{e}_{\rho} + \frac{\cosh{\xi}\cos{\eta}-1}{\cosh{\xi}-\cos{\eta}}\mathbf{e}_z.
			\end{align}
\end{subequations}
			We take the torus $\mathcal{S}$ to be the surface $\xi =
\xi_0$, with aspect ratio $s_0 = \cosh{\xi_0} = b / a$, then the unit normal
pointing into the fluid $\mathbf{n} = - \mathbf{e}_{\xi_0}$, and spatial
infinity corresponds to $\xi, \eta \to 0$. Note that we can relate the toroidal
coordinates to the poloidal angle $\theta \in [0,2\pi)$ on the surface of the
torus; a point on the torus' surface is specified by $\rho = b + a\cos{\theta}$
and $z  = a \sin{\theta}$, so that $\theta$ is related to $\xi, \eta$ through~(\ref{eq:9}) and 
			\begin{equation}
				\label{eq:14} \tan{\theta} = \frac{z}{\rho - b}\cdot
			\end{equation}

		\subsection{\label{subsec:2.3}Flow field around a phoretic torus}
		
			The tori considered in this work have axisymmetric
surface chemistry, and hence the nondimensional surface activity and surface
mobility $A = A(\eta)$ and $M = M(\eta)$ respectively are solely functions of
$\eta$.
			
			\subsubsection{\label{subsubsec:2.3.1} Product concentration}
			In the case of fixed-flux, we can think equivalently in
terms of solute reducing at the boundary, or product increasing at the boundary,
the solutions differing by a minus sign.
Here we work in terms of positive product concentrations relative to a zero
concentration at infinity.
				The axisymmetric activity gives rise to $c =
c(\xi, \eta)$, and hence Laplace's equation,(\ref{eq:2}), becomes
				\begin{equation}
					\label{eq:15} 0 = \frac{\partial}{\partial \xi} \left(\frac{\sinh{\xi}}{\cosh{\xi}-\cos{\eta}} \frac{\partial c}{\partial \xi} \right) + \frac{\partial}{\partial \eta} \left(\frac{\sinh{\xi}}{\cosh{\xi}-\cos{\eta}} \frac{\partial c}{\partial \eta} \right).
				\end{equation}
				Let $s = \cosh{\xi}$, $t = \cos{\eta}$, and let $P_{m-1/2}, Q_{m-1/2}$ denote Legendre functions of the first and second kind of degree $m - 1/2$ respectively. Then eq.~(\ref{eq:15}) admits a solution, which decays at spatial infinity and is $2 \pi$ periodic in $\eta$, of the form~\cite{PM1}
				\begin{equation}
					\label{eq:16} c(\xi, \eta) = \sqrt{s - t} \sum_{n=0}^{\infty}{\vphantom{\sum}}'\left(A_n \cos{n \eta} + B_n \sin{n \eta} \right) P_{n - 1/2}(s),
				\end{equation}
				where $\sum_{n=0}^{\infty}{\vphantom{\sum}}'$ indicates that the term with $n=0$ is multiplied by $1/2$. Imposing the Neumann boundary condition~(\ref{eq:1}) then requires
				\begin{equation}
					\label{eq:17} \mathbf{e}_{\xi} \cdot \bm{\nabla}c = \frac{1}{h_{\xi}}\frac{\partial c}{\partial \xi} = A(\eta) \quad \text{ on } \xi = \xi_0.
				\end{equation}
				Using the form of the concentration~(\ref{eq:16}), we can rearrange the above to give
\begin{subequations}
				\begin{align}
					\label{eq:18}
\frac{A(\eta)d}{\sinh{\xi_0}\sqrt{s_0 - t}} &=
\sum_{n=0}^{\infty}{\vphantom{\sum}}' \left(A_n \cos{n \eta} + B_n \sin{n \eta}
\right)\left( \tfrac{1}{2}P_{n-1/2}(s_0) + (s_0 - t) P'_{n-1/2}(s_0)\right), \\
					\label{eq:19} &=
\sum_{n=0}^{\infty}{\vphantom{\sum}}' \left(C_n \cos{n \eta} + D_n \sin{n \eta}
\right).
				\end{align}
\end{subequations}
				The Fourier coefficients $(C_n, D_n)$ can be related to the $(A_i, B_i)$ through standard trigonometric identities. The coefficients $A_n$ and $B_n$ can be calculated numerically after truncating the system.
				
			\subsubsection{\label{subsubsec:2.3.2}Slip velocity}

				Once the concentration has been calculated, the slip velocity can be obtained from eqs.~(\ref{eq:4}) and~(\ref{eq:16}) in terms of the nondimensional surface mobility $M$ as
				\begin{align}
					\nonumber \mathbf{u}_s &= \frac{M}{a} \frac{\partial c}{\partial \theta} \mathbf{e}_{\theta} =
\frac{M}{h_{\eta}}\frac{\partial c}{\partial \eta} \mathbf{e}_{\eta}, \\
					\label{eq:20} &= \frac{M}{d} \Big[ \tfrac{1}{2} c(\xi_0, \eta) \sin{\eta} + \left(s_0 - \cos{\eta}\right)^{3/2} \sum_{n=1}^{\infty} \left( -n A_n \sin{n \eta} + n B_n \cos{n \eta} \right) P_{n-1/2}(s_0) \Big]\mathbf{e}_{\eta}.
				\end{align}
				As the $A_n$ are linear in $d$, the slip
velocity depends on the geometry of the system only through the aspect ratio
$s_0$ and not the absolute size of the torus, in an analagous manner to the size
independent swimming velocity of spherical Janus particles in the fixed flux
limit~\cite{RG1}.
				
			\subsubsection{\label{subsubsec:2.3.3}Stokes streamfunction}
			
				The Stokes flow problem for a torus has been
solved by Pell and Payne~\cite{WHP1} for a fixed torus in uniform flow, and by
Leshansky and Kenneth~\cite{AML1} for a torus with slip velocity symmetric about
$\eta = \pi$. In this section, we summarize the results and apply them to the
case that the slip velocity is given by eq.~(\ref{eq:20}). To identically solve
the continuity equation~(\ref{eq:6}), we introduce the vector potential
$\mathbf{A}$ such that $\mathbf{u} = \bm{\nabla} \times \mathbf{A}$.  For
axisymmetric flows we may take $\mathbf{A} = \Psi\mathbf{e}_{\phi}/\rho$, and then the velocity components are given, in cylindrical and toroidal coordinates respectively,
\begin{subequations}
				\begin{alignat}{2}
					\label{eq:21} u_{\rho} &=
-\frac{1}{\rho}\frac{\partial \Psi}{\partial z},\quad &u_{z} &=\frac{1}{\rho}\frac{\partial \Psi}{\partial \rho}, \\
					\label{eq:22} u_{\xi} &= \frac{1}{\rho
h_{\eta}}\frac{\partial \Psi}{\partial \eta},\quad &u_{\eta} &= -\frac{1}{\rho h_{\xi}}\frac{\partial \Psi}{\partial \xi}\cdot
				\end{alignat}
\end{subequations}
				The stream function $\Psi$ satisfies $L_{-1}^2 \Psi = 0$, where the operator $L_k$ is given by
				\begin{equation}
					\label{eq:23} L_k = \frac{\partial^2}{\partial \rho^2} + \frac{\partial^2}{\partial z^2} + \frac{k}{\rho}\frac{\partial}{\partial \rho}\cdot
				\end{equation}
				{For the boundary conditions on the streamfunction, we ask that the surface $\xi = \xi_0$ be a streamline $\Psi = \chi$, where $\chi$ is a constant, and that on the surface there is a slip velocity $\mathbf{u} = u_s(\eta)\mathbf{e}_\eta$ given by eq.~(\ref{eq:22}).  Working in the reference frame where the torus is fixed, we require the background flow at infinity to be uniform with $\Psi \sim -\frac{1}{2}U \rho^2$ as $r \to \infty$, where $U$ is the swimming velocity of the torus. Writing $\Psi = -\frac{1}{2}U\rho^2 + \psi + \chi \varphi$, the boundary conditions become, as in~\cite{AML1},}
\begin{subequations}
				\begin{alignat}{2}
					\label{eq:24}\psi &= \tfrac{1}{2}U
\rho^2,\quad
					&\frac{\partial \psi}{\partial \chi} &= U \rho \frac{\partial \rho}{\partial \xi} -\rho h_{\xi}u_s(\eta), \\
					\label{eq:25}\varphi &= 1,\quad
					&\frac{\partial \varphi}{\partial \xi} &= 0,
				\end{alignat}
\end{subequations}
				on the surface $\xi = \xi_0$. Setting $\psi^k$
to be
solutions to $L_k \psi^k = 0$, we can represent $\psi = \tfrac{1}{2}
\rho^2 \left(\psi^1 + (r^2 + d^2)\psi^3 \right)$~\cite{WHP1, LEP1}. We can then
solve for $\psi^3$ and $\psi^1$, which decay at infinity and are $2 \pi$
periodic in $\eta$, using separation of variables to get
				\begin{equation}
					\label{eq:26} \psi = \tfrac{1}{2}\rho^2 (s-t)^{1/2} \sum_{n=0}^{\infty}{\vphantom{\sum}}' \left( W^1_n(s) \cos{n \eta} + W^2_n(s) \sin{n \eta}\right),
				\end{equation}
				where the coefficient functions $W_m^i$ are
\begin{subequations}
				\begin{align}
					\label{eq:27} W^1_n(s) &= a_n P_{n-1/2}(s) + c_n s P'_{n-1/2}, \\
					\label{eq:28} W^2_n(s) &= b_n P_{n-1/2}(s) + d_n s P'_{n-1/2}.
				\end{align}
\end{subequations}
				Making use of Heine's identity~\cite{EH1}, 
				\begin{equation}
				\label{eq:29} \frac{1}{\sqrt{s - \cos{\eta}}} = \frac{\sqrt{2}}{\pi} Q_{-1/2}(s) + \frac{2\sqrt{2}}{\pi} \sum_{n=1}^{\infty} Q_{n-1/2}(s) \cos{n \eta},
				\end{equation}
				the boundary conditions~(\ref{eq:24}) become
\begin{subequations}
				\begin{alignat}{2}
					\label{eq:30} W_n^1(s_0) &=
\frac{2\sqrt{2}}{\pi}U Q_{n-1/2}(s_0),\quad & \frac{dW^1_n}{d s_0} &=  U \frac{2\sqrt{2}}{\pi} Q'_{n-1/2}(s_0) - \frac{2 E_n}{\sinh^2{\xi_0}},\\
					\label{eq:31} W_n^2(s_0) &= 0,\quad & \frac{dW^2_n}{d s_0} &= - \frac{2F_n}{\sinh^2{\xi_0}},
				\end{alignat}
\end{subequations}
				where the $E_n$ and $F_n$ are the cosine and sine Fourier coefficients of $u_s / \left( s_0 - t \right)^{1/2}$, which can be obtained from the expression for the slip velocity~(\ref{eq:20}). The boundary conditions for $\varphi$~(\ref{eq:25}) are exactly as in~\cite{AML1,WHP1} (given as $\psi_1$ in the references), whence 
				\begin{equation}
					\label{eq:32} \varphi = \frac{\rho^2}{d^2}(s-t)^{1/2} \sum_{n=0}^{\infty}{\vphantom{\sum}}' \left(e_n P_{n-1/2}(s) + f_nsP'_{n-1/2}(s)\right) \cos{n \eta},
				\end{equation}
				with the coefficients $f_n$ and $e_n$ given by
\begin{subequations}
				\begin{align}
				\label{eq:33} & f_n s_0 P'_{n-1/2}(s_0) +
e_nP_{n-1/2}(s_0) = \frac{3}{\pi \sqrt{2}} Q^{-2}_{n-1/2}(s_0), \\
				\label{eq:34} & f_n \frac{d}{d s_0}\left(s_0 P'_{n-1/2}(s_0)\right) + e_nP'_{n-1/2}(s_0) = \frac{3}{\pi \sqrt{2}} \frac{d}{d s_0} \left( Q^{-2}_{n-1/2}(s_0) \right).
				\end{align}
\end{subequations}
				{Following Leshansky and Kenneth~\cite{AML1}}, linearity allows us to split
				$\psi$ into the contributions from the uniform background flow and the slip
				velocity, $\psi = U\psi^{(g)} + \psi^{(p)}$, {where $U$ is the undetermined non-dimensional swimming velocity}. {Here $\psi^{(p)}$ is the streamfunction appropriate for a fixed torus with slip velocity $u_s$, and $\psi^{(g)}$ solves the problem of a torus with no slip velocity in a background flow with $U = 1$.} This has the effect that $a_n = U a_n^{(g)} +
				a_n^{(p)}$ and similarly for other coefficients. The coefficients $b_n, d_n$ depend only on the slip velocity and hence do not pick up contributions from the background flow. The constant $\chi$ is found from the requirement that the pressure is single valued, see~\cite{AML1,WHP1,LEP1} for details, as
				\begin{equation}
					\label{eq:35} \chi =
-\frac{d^2\sum_{n=0}^{\infty}{\vphantom{\sum}}' {(Uc^{(g)}_n + c^{(p)}_n)}}{2
\sum_{n=0}^{\infty}{\vphantom{\sum}}' f_n} { \equiv - \frac{d^2}{2}(U\alpha + \beta)}.
				\end{equation}
				Finally, to determine $U$, we impose that
the force on the torus must vanish. We find the force  on the torus, which by symmetry is along the $z$ axis, from
				\begin{equation}
					\label{eq:36} \frac{F}{8 \pi \mu} =
\lim_{r \to \infty} \frac{r(\psi + \chi \varphi)}{\rho^2},
				\end{equation}
				which is equivalent to evaluating the fraction
at $\eta = \xi = 0$. Using the split form of $\psi$ and setting $F = 0$
in~(\ref{eq:36}), we can derive the expression for the swimming velocity of the torus as in~\cite{AML1}
				\begin{equation}
					\label{eq:37} U = - \frac{\sum_{n=0}^{\infty}{\vphantom{\sum}}' \left( \left(n^2 - \tfrac{1}{4}\right) \left( c_n^{(p)} - \beta f_n \right) + 2\left( a_n^{(p)} - \beta e_n\right)  \right)}{\sum_{n=0}^{\infty}{\vphantom{\sum}}'\left( \left(n^2 - \tfrac{1}{4}\right) \left( c_n^{(g)} - \alpha f_n \right) + 2\left( a_n^{(g)} - \alpha e_n\right)  \right)}\cdot
				\end{equation}
				We now apply the solution to analyse special cases of particular interest.

	\section{\label{sec:3} Results}
				
		\subsection{\label{subsec:3.1} The optimal uniform toroidal pump}
			
\begin{figure}[t]
				\includegraphics{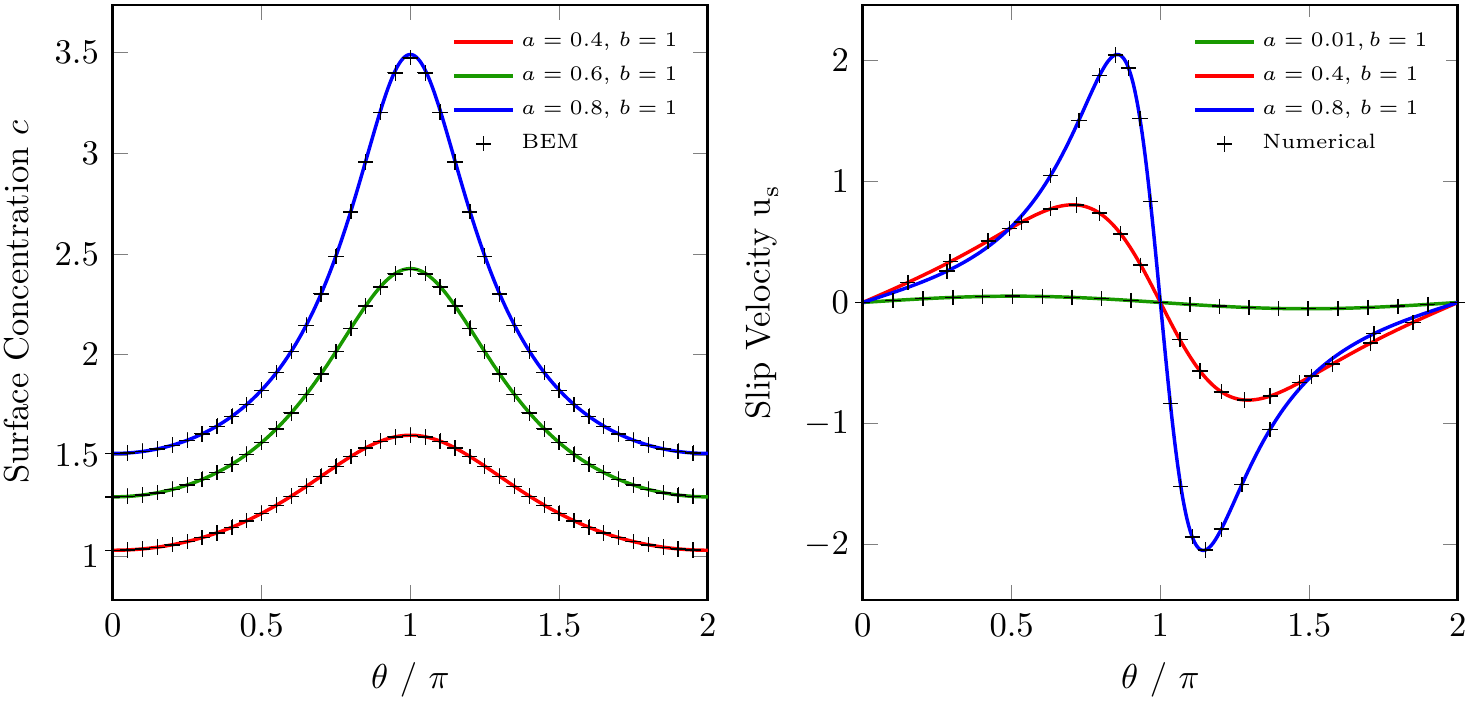}
				\caption{\label{fig:1} {Validation
of the series solution.} Left: Surface
concentration as a function of the poloidal angle $\theta$, with the analytical series
solution (lines) compared against a numerical solution (crosses) computed with a
3D boundary element method~\cite{TDMJ1}. {} Right: Slip velocity as a function of
$\theta$, with the series
solution (lines) compared against results obtained from numerically
differentiating the validated series solution for the concentration (crosses).}
			\end{figure}

			First, we consider the chemically-uniform case where both activity
$A$ and mobility $M$ are constants, $\pm 1$. We can use Heine's identity~(\ref{eq:29}) in the expressions for the concentration coefficients~(\ref{eq:18}) to set $B_n = D_n = 0$, and then the $A_n$ are determined through
\begin{subequations}
			\begin{align}
				\label{eq:38} \frac{2\sqrt{2}}{\pi}
Q_{-1/2}({s_0})
= A_0 &\left[ \tfrac{1}{2} P_{-1/2}(s_0) + s_0 P'_{-1/2}(s_0) \right] - A_1 P'_{1/2}(s_0), \\
				\nonumber \label{eq:39} \frac{2\sqrt{2}}{\pi}
Q_{n-1/2}({s_0}) = A_n &\left[ \tfrac{1}{2}P_{n-1/2}(s_0) + s_0 P'_{n-1/2}(s_0)\right] \\
				 - \tfrac{1}{2} &\left[ A_{n+1}P'_{n+1/2}(s_0) + A_{n-1}P'_{n-3/2}(s_0)\right].
			\end{align}
\end{subequations}
			Truncating the system~(\ref{eq:38}, \ref{eq:39}) at an
appropriately chosen $n$ enables numerical computation of the $A_i$. Throughout
this study we take between $n = 45$ and $n = 5$ coefficients for aspect ratios in the range $s_0 = 1.1$ and $s_0 = 30$, similar to the numbers used by Leshansky and Kenneth~\cite{AML1}. To check the numerical accuracy, we compare the surface concentration calculated from the analytical solution against the surface concentration obtained from a 3D regularized boundary element method~\cite{TDMJ1}. The comparison (Fig.~\ref{fig:1}, left) shows the surface concentrations to be in good agreement with a maximal percentage difference of less than $0.5\%$ for $a = 0.8$ and $b = 1$.		
			
\begin{figure}[t]
				\includegraphics{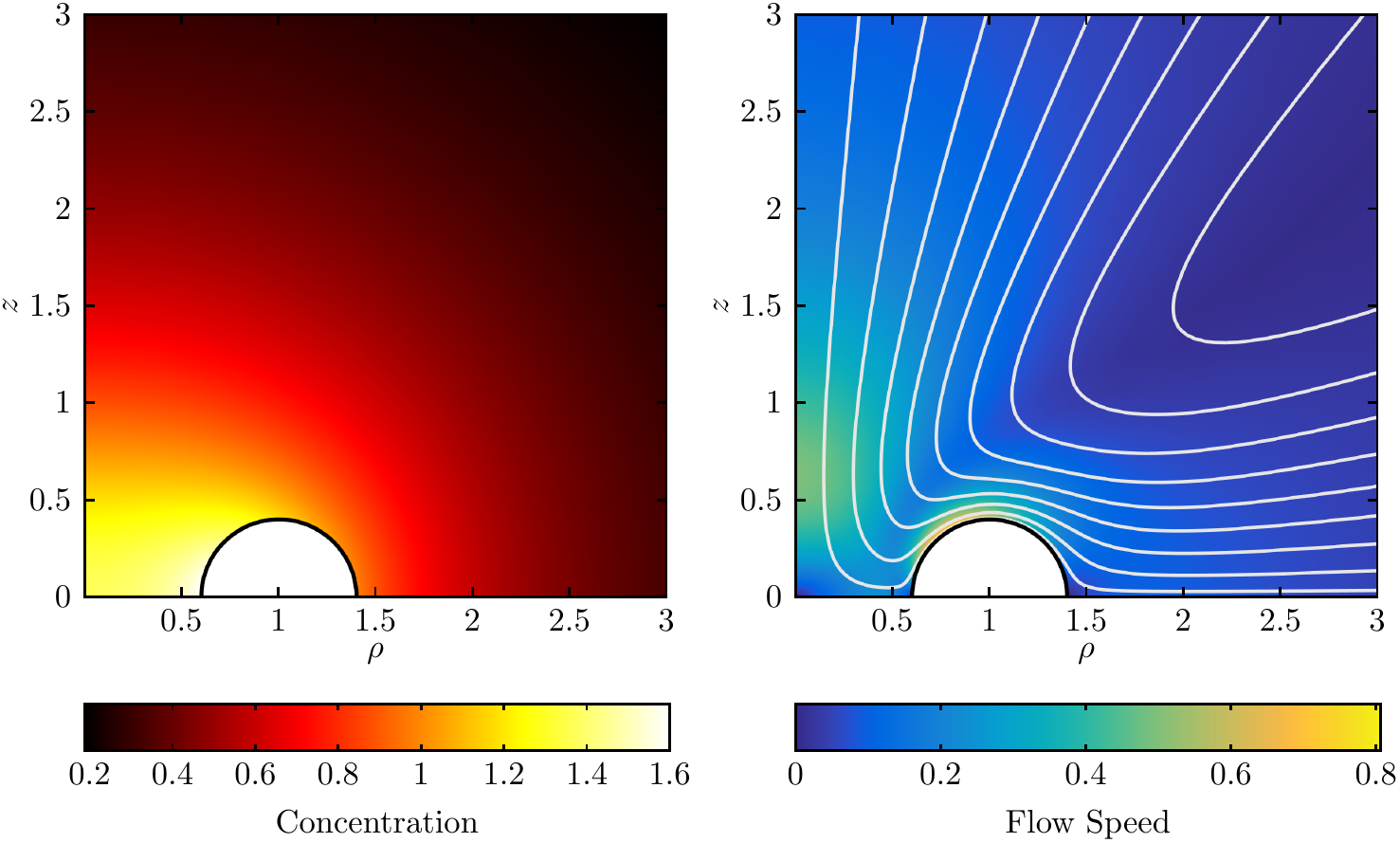}
				\caption{\label{fig:2}Left: Product
concentration around { uniform} torus for $a=0.4$, $b=1$ and constant surface activity $A =
+1$, computed with the analytical series solution. Right: Flow speed around
torus with $a = 0.4$, $b = 1$ and $A = M = +1$, computed with the analytical
series solution; flow streamlines shown in white.}
			\end{figure}

			In contrast to isolated spherical particles, where a
uniform surface chemistry is unable to drive flow~\cite{RG1,SM1}, a uniform
phoretic torus generates a pumping flow field. When the torus releases product,
confinement causes this product to build-up in the central hole (see
Figs.~\ref{fig:1}~and~\ref{fig:2}, left), and the resulting concentration
gradients give rise to the slip velocities on the surface (see Fig.~\ref{fig:1},
right) which drive the overall flow. The confining effect becomes more
pronounced in the limit where the central hole shrinks and the aspect ratio $s_0
= b/a \to 1$. In this limit, the concentration becomes increasingly localized in
the central hole, leading to large peaks in the slip velocity moving closer to
either side of $\theta = \pi$ as the differences in the confining effect between
neighbouring points becomes greater. 

However, on the outer edge of the torus, the largest slip velocities do not
occur in the limit where the central hole vanishes (Fig.~\ref{fig:1}, right).
Since the strength of pumping will depend in some sense on an integral of slip
velocities over the toroidal surface, this result suggests that for fixed
surface area, there may be a non-trivial optimal pump.

			We test this notion by calculating the flow in the bulk of the fluid via the streamfunction $\Psi$. For uniform surface chemistry, symmetry prevents the system from swimming, thus $U = 0$. Additionally we require that $\Psi$ is antisymmetric in $\eta$ around $\eta = \pi$, which forces $\chi = 0$, and hence $\Psi = \psi$ with $W_n^1(s) = 0$. The Fourier coefficients $F_n$ in~(\ref{eq:31}) can be found from the slip velocity~(\ref{eq:20}) in terms of the concentration coefficients $A_i$ as
			\begin{equation}
				\label{eq:40} F_n = \frac{M}{d} \left(
-nA_ns_0P_{n-1/2}(s_0) + \left(\tfrac{n}{2} -
\tfrac{1}{4}\right)A_{n-1}P_{n-3/2}(s_0)+ \left(\tfrac{n}{2} +
\tfrac{1}{4}\right) A_{n+1}P_{n+1/2}(s_0) \right).
			\end{equation}
			Using equations~(\ref{eq:22}), these coefficients then allow us to compute the flow. The right of Fig.~\ref{fig:2} shows the flow speed $u = (u_{\xi}^2 + u_{\eta}^2)^{1/2}$ and corresponding streamlines. When $AM = 1$, the fluid comes in along the $\rho$ direction and is pushed out again along the $z$ direction (if $AM = -1$ the flow direction is reversed). 
				
			By symmetry the pumping torus is force-free, thus we
expect the far-field
flow to behave like a stresslet, i.e.~the solution to Stokes flow driven
by a point stress~\cite{GKB1}. As such, we predict flow decays like $u \propto 1/r^2$ away
from the torus. Considering the flow speed along the $\rho$ axis, we should
therefore see
			\begin{equation}
				\label{eq:41} |u|_{z = 0} \sim \frac{k}{\rho^2}\cdot
			\end{equation}
			Along $z = 0$, we have $\eta = 0$ and $|\mathbf{u}| = |u_{\xi}|$. From eqs.~(\ref{eq:26}) and~(\ref{eq:22}) we find that along $\eta = 0$
			\begin{equation}
				\label{eq:42} u_{\xi} = \frac{1}{2} \sinh{\xi} (s-1)^{1/2} \sum_{n=1}^{\infty} n W_n^2(s).
			\end{equation}
			Using $\sinh{\xi} = \sqrt{(s-1)(s+1)}$, $(s - 1) \sim 2\frac{d^2}{\rho^2}$ along $\eta = 0$ as $s \to 1$, and the asymptotic forms of the toroidal harmonics~\cite{MA1} as $s \to 1$ in~(\ref{eq:42}) we find that 
			\begin{equation}
				\label{eq:43} k = \Bigl| \sqrt{2}d^2 \sum_{n=1}^{\infty} n\left(b_n + \tfrac{1}{2}d_n (n^2 - \tfrac{1}{4})\right)\Bigr|,
			\end{equation}
			giving the strength of the pumping $k$ for a given torus in terms of the stream function coefficients $b_n$ and $d_n$. We can use this expression to optimize the torus to maximize the pumping.
			\begin{figure}[t]
				\includegraphics{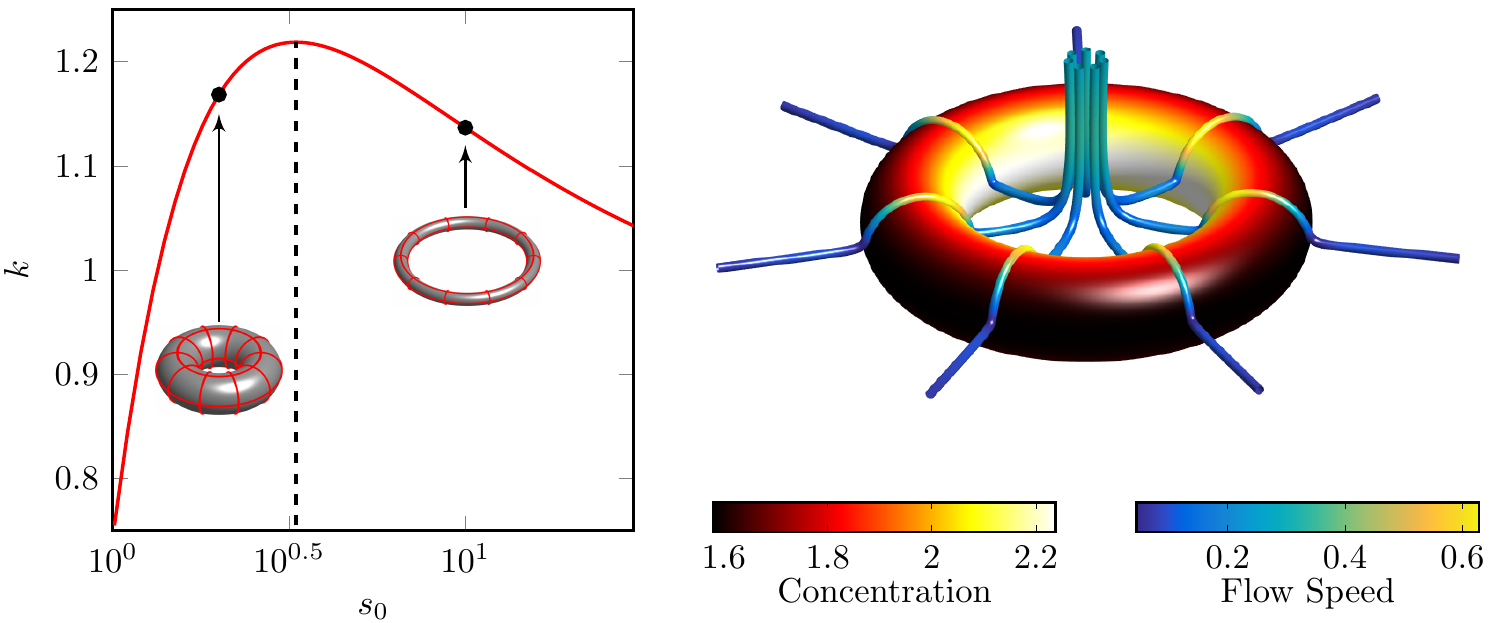}    
				\caption{\label{fig:3} Left: Decay coefficient
$k$ as a function of aspect ratio $s_0$ for fixed surface area $A_s = 4\pi^2$,
as calculated by the series~\eqref{eq:43}. Tori with $s_0 = 2$
and $s_0 = 10$ are shown for comparison. Right: The concentration and flow
streamlines of the optimal uniform torus with $A_s = 4 \pi^2$, as calculated
with the analytical series solution.}
			\end{figure}					
	
			The area of the active surface on the torus
governs the net flux of product into the solution~\cite{TDMJ2}. Fixing the
surface area $A_s = 4 \pi^2 a b$, we still have the freedom to vary the aspect
ratio of the torus $s_0 = b/a = \cosh{\xi_0}$. Measuring the pumping strength of
the uniform torus through $k(s_0)$ of (\ref{eq:41}), we seek the optimal aspect
ratio which maximizes the pumping. From the left of Fig.~\ref{fig:3}, we see
that there is a peak in the stresslet strength with an optimal aspect ratio at
$s_0 = 3.31$. The corresponding optimal torus is shown on the right of
Fig.~\ref{fig:3}, notice in particular that the optimum interestingly occurs for
a torus with a central hole, so that pumping is not maximized for the aspect
ratio which maximizes geometric confinement.
	
		\subsection{\label{subsec:3.2} Optimal {smooth-activity} Janus torus}
	
\begin{figure}[t]
				\includegraphics{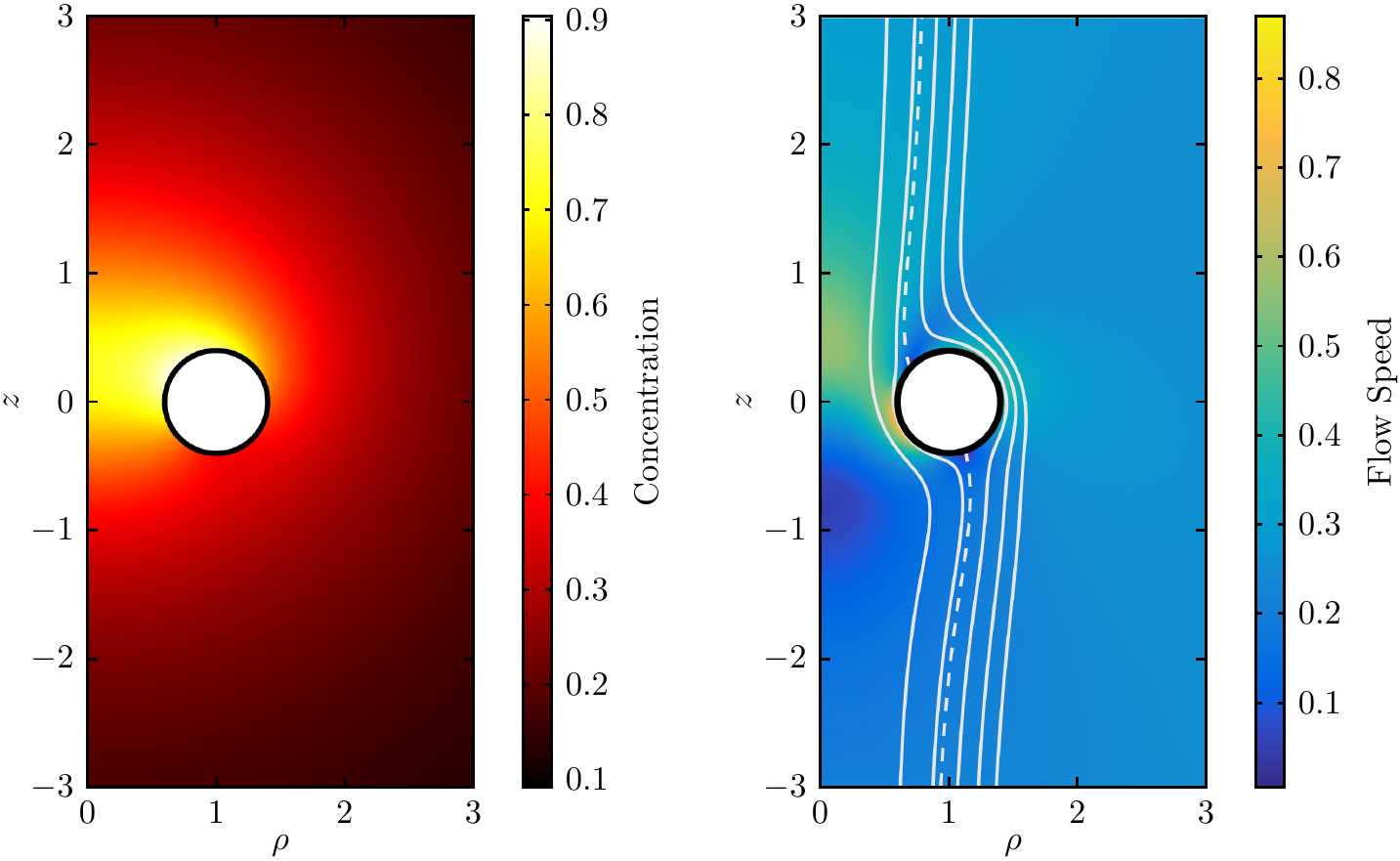}
				\caption{\label{fig:4} Left: Concentration
around {the smooth-activity} Janus torus with $a = 0.4$ and $b=1$ computed from the analytical series solution. Right: Flow speed around {the smooth-activity} Janus torus with $a = 0.4$ and $b=1$. Streamlines in the body frame in white, with
dashed streamlines correspond to dividing streamlines $\psi = \chi$.}
			\end{figure}

We consider next a smoothed approximation to a Janus torus. We take the
{differentiable surface} activity of the form
			\begin{equation}
			\label{eq:44} A = \tfrac{1}{2}\left(1+\sin{\theta} \right) = \tfrac{1}{2}\left( 1 +  \frac{\sin{\eta}\sinh{\xi_0}}{\cosh{\xi_0} - \cos{\eta}}\right),
			\end{equation}
while the surface mobility is kept constant $M = \pm
1$~\citep{nourhani2016geometrical}. Such {smooth activity functions
are numerically convenient, and may be} employed~\cite{YI1} for studies of
spherical Janus particles, {which in reality have discontinuities in
the surface activity at the boundary of the active and inert portions of their
surfaces}. 
The top-bottom asymmetry in the surface activity produces
asymmetric concentration gradients and slip velocities (see Fig.~\ref{fig:4})
which allow the torus to have a nonzero swimming velocity. For $M = 1$, 
swimming  occurs along the negative $z$ direction ($M= -1$ reverses
flow and swimming directions).

			The confinement and changes in the surface activity lead
to two areas on the surface of the torus where there are significant slip
velocities, visible on the right of Fig.~\ref{fig:4}.  The first area of large
slip velocities occurs on lower half of the central hole, due to the decrease in
the value of the activity, while the second occurs towards the outer edge of
the torus as a result of the increased freedom for the product to diffuse. As
before we validate the accuracy of these solutions by comparing the results
obtained by truncating the analytical expressions against the boundary element
method, the details of which are described in Ref.~\cite{TDMJ1}. Over the range
$1.1 \leq s_0 \leq 10$, with fixed $A_s = 4 \pi^2$, we obtain percentage
differences between the concentrations calculated using the two methods of at
most than $1.1 \%$ for $s_0 = 1.1$.  However, this error is likely attributable
to the difficulty of modelling small aspect ratio tori accurately with the
boundary element method; larger aspect ratios, which are easier to model with
boundary elements, give smaller percentage differences; for example, for $s_0 = 2.81$ the
difference calculated was less than $0.03 \%$.
			
\begin{figure}[t]
				\includegraphics{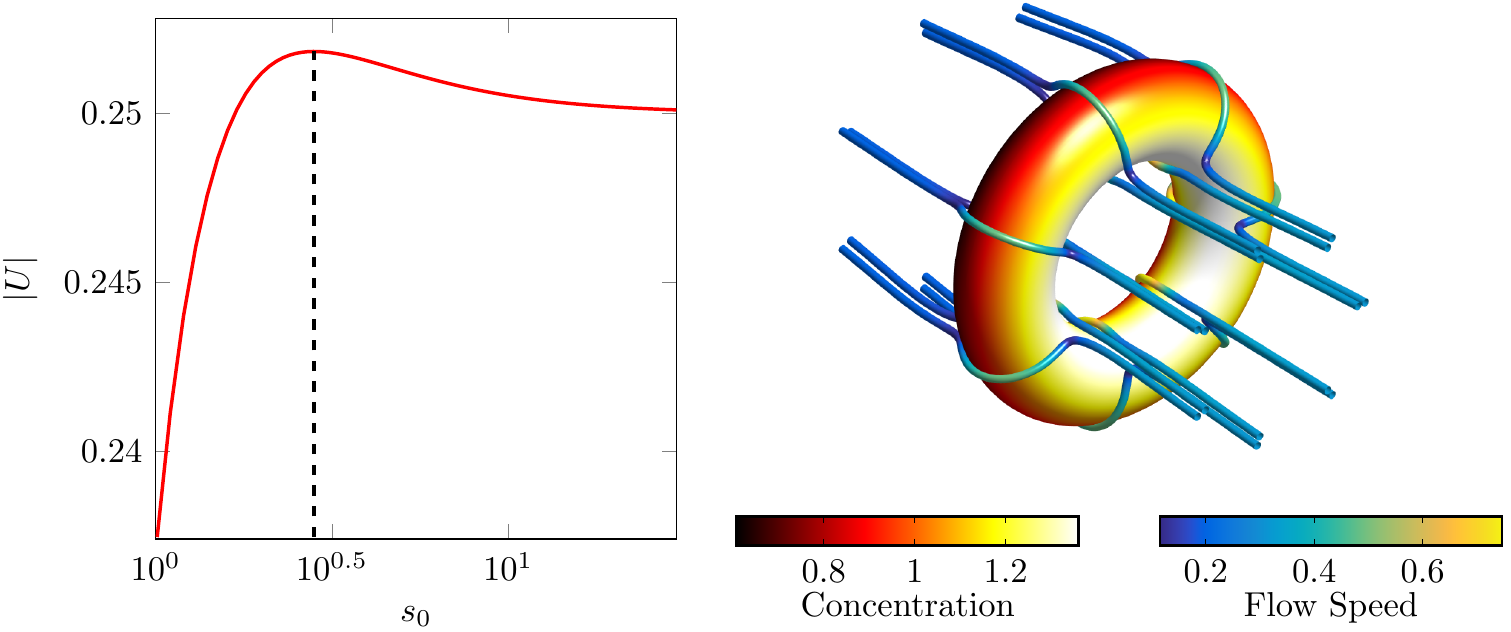}
				\caption{\label{fig:5} Left: Swimming speed
$|U|$ of {the smooth-activity} Janus torus as a function of $s_0$. Right:
The optimal {smooth-activity} Janus torus with surface concentration and streamlines in the body
frame. Computed with the analytical series solution.}
			\end{figure}

			As in \S\ref{subsec:3.2}, we compute $|U|$ as we vary the aspect ratio $s_0$ (Fig.~\ref{fig:5}, left). There is a peak at $s_0 = 2.81$, with swimming speed $|U| = 0.2518$, before the swimming velocity appears to tend towards an asymptote for larger $s_0$. In the large $s_0$ limit, there is no self-interaction between the opposing ends of the torus. The swimming speed being independent of the torus shape in this limit is similar to the result for spherical Janus particles, whose swimming speed is independent of their size in this classical phoretic framework~\cite{RG1}.
			
The optimal {smooth-activity} torus, shown on the right of Fig.~\ref{fig:5}, has a
smaller aspect ratio $s_0 = 2.81$ than the optimal toroidal pump $s_0 = 3.31$.
This difference arises because the swimming optimum arises from the balance of
two physical effects; maximising surface chemical gradients/slip flow, as with
the pump, and minimising drag via hydrodynamic interactions. The proximity
of the opposing side of the torus reduces drag by pulling its neighbouring
segments along,
an effect which is also observed for groups of sedimenting
spheres~\cite{KOLFJ1}. Thus, the swimming optimum has a smaller central hole
than the pump; suboptimal slip generation is balanced by reduced drag in this
fatter torus.
			
		\subsection{\label{subsec:3.3} Optimal Janus torus}
	
			Finally, we consider a Janus torus, with both the surface activity and surface mobility having a step-discontinuity at $\theta = \pi$, given by 
			\begin{equation}
				\label{eq:45} A = M = 
				\begin{cases}
					1 \quad \text{if $\theta \in (0,\pi)$}, \\
					0 \quad \text{if $\theta \in (\pi, 2\pi)$}.
				\end{cases}
			\end{equation}
			
			The set of linear equations for the concentration
coefficients~(\ref{eq:16}) becomes ill conditioned when attempting to solve for
a large number of coefficients. This makes it difficult to accurately resolve
the step discontinuities in the activity and mobility using the series solution.
However, due to the dissipative nature of Stokes flow, the wiggles in the
boundary condition are smoothed out quickly in the fluid, and the swimming
velocity is dependent upon an integral of the boundary slip. Thus we expect the
error introduced by the low number of coefficients to be small, and we check the error introduced by comparing the computed results against the boundary element method~\cite{TDMJ1}. The swimming speeds calculated using the two methods have percentage differences of at most $3 \%$.

\begin{figure}[t]
				\includegraphics{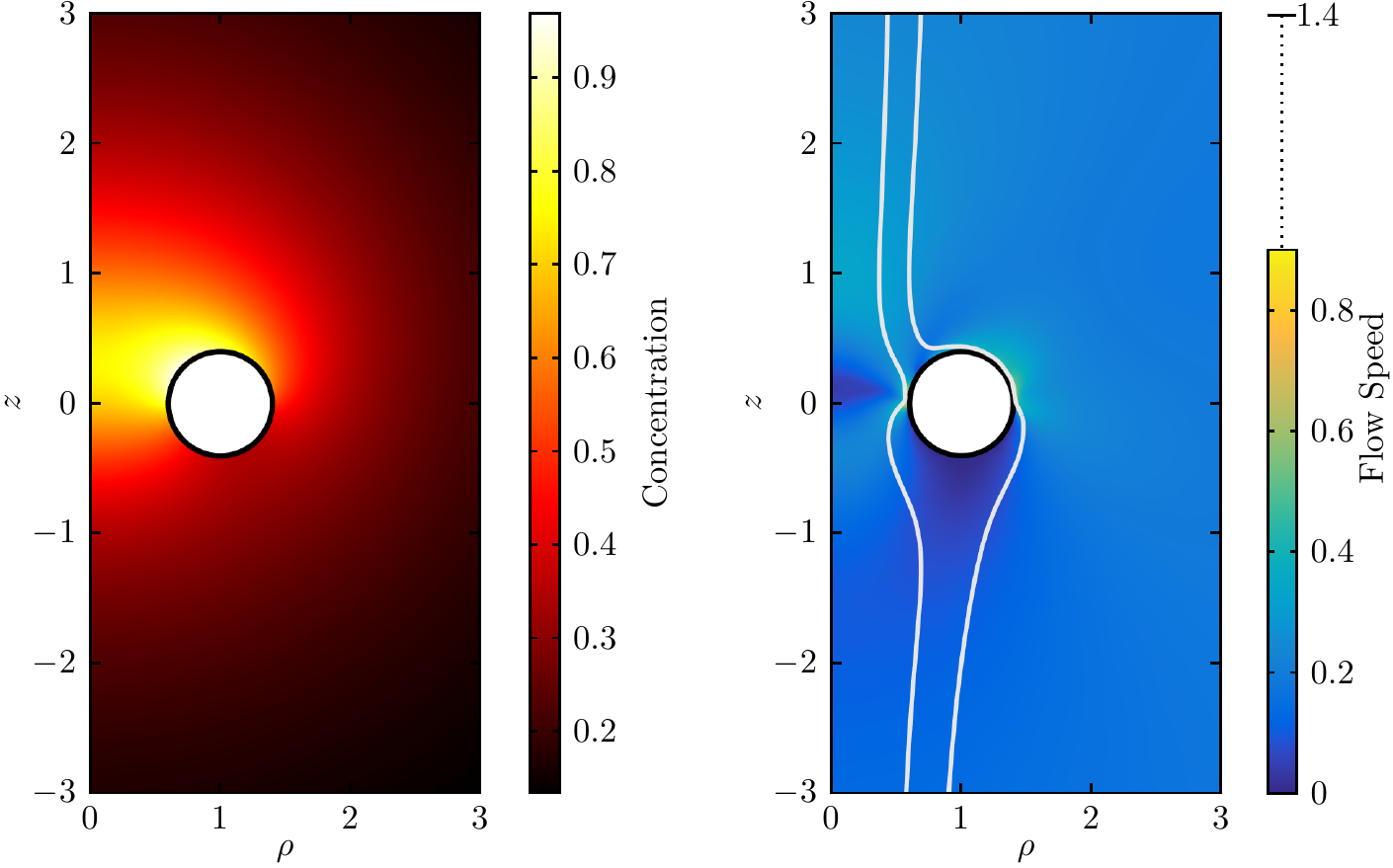}
				\caption{\label{fig:6} Left: Concentration
around a Janus torus with $a = 0.4$ and $b=1$. Right: Flow speed around Janus
torus with $a = 0.4$ and $b=1$. Computed with the analytical series solution.}
			\end{figure}

\begin{figure}[t]
				\includegraphics{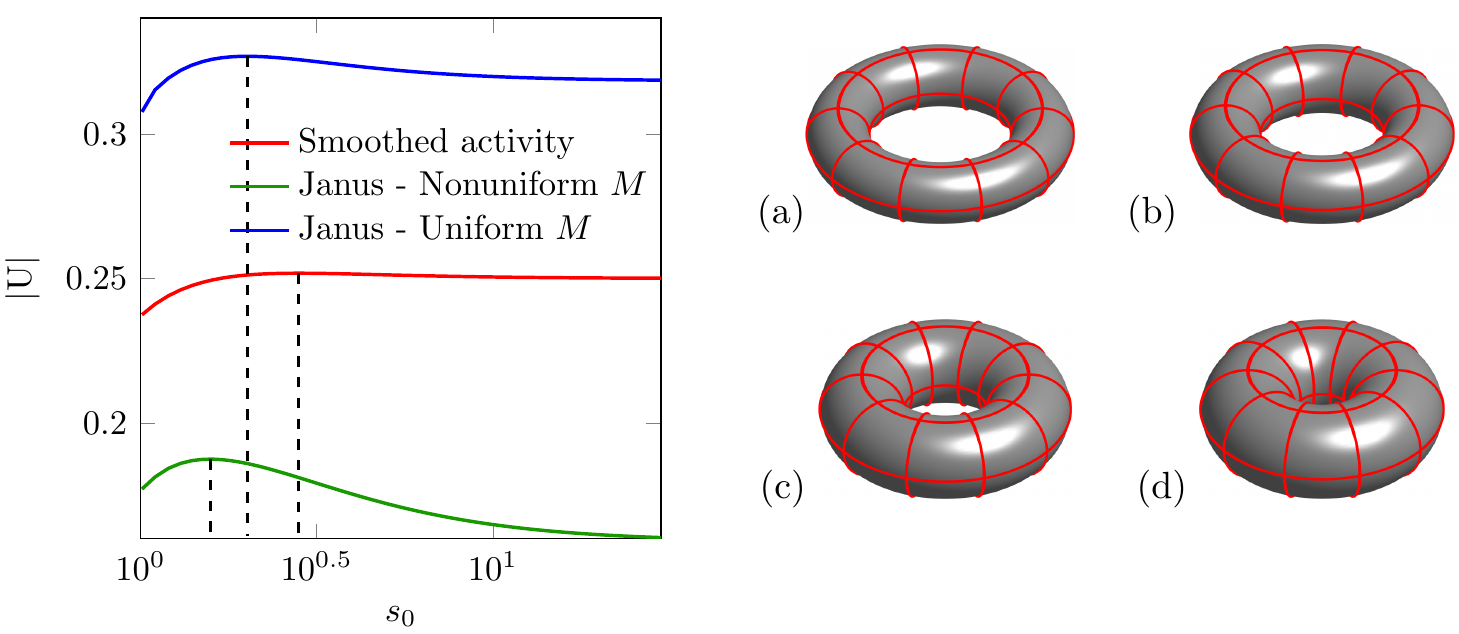}
				\caption{\label{fig:7} Left: Swimming speed
$|U|$ of Janus torus with surface chemistry given by Eq.~(\ref{eq:45}), computed
with the analytical series solution.  Right:
Plots of the optimal tori: (a) Uniform pump, (b) {smooth-activity} Janus torus, (c)
uniform mobility Janus torus, (d) nonuniform mobility Janus torus.}
			\end{figure}

The concentration around the Janus torus
(Fig.~\ref{fig:6}, left) is similar to concentration found for the first-mode
approximation in \S\ref{subsec:3.2}, with the product slightly more evenly
distributed around the {smooth-activity} torus. These differences can produce larger
changes in the slip velocity, and thus in the swimming speed. Away from the
surface of the torus the flow fields (Fig.~\ref{fig:6}, right) calculated using
the two methods described are in good agreement. Notice that the colorbar has
been truncated at $u = 0.9$, highlighting the structure of the flow field. The
most prominent difference, compared with the previously considered torus, is
that the area of large surface slip velocities on the lower half of the central hole
in Fig.~\ref{fig:4} no longer appears for the Janus torus with discontinuous
mobility.						
			We again aim to determine the optimal aspect ratio which maximizes the swimming speed of the Janus torus. The green line on Fig.~\ref{fig:7} shows the swimming speed of the Janus torus with nonuniform mobility~(\ref{eq:45}). There is a pronounced peak at $s_0 = 1.58$, which corresponds to a torus with a smaller central hole than found for the first-mode {smooth-activity Janus} torus, shown as the red line. 

To determine whether this difference is solely due to the change in mobility or
whether it is also affected by the surface activity, we repeat the calculation
for a Janus torus with uniform mobility $M=1$, shown as the blue line on
Fig.~\ref{fig:7}. Whilst perhaps not experimentally relevant, the uniform
mobility case has been considered in previous
studies~\cite{uspal2015self,nourhani2016geometrical} and provides an indication
of the behaviour of tori with intermediate non-zero mobilities on the inert cap.
For the Janus torus with uniform mobility, the optimal swimming speed is
achieved for $s_0 = 2.01$, for a swimming speed $|U| = 0.3268$, greater than for
the {smooth-activity} Janus torus.  Additionally, the peak in the
swimming velocity is much sharper than for the first-mode {smooth-activity} Janus torus.  Notably, in the fixed-flux limit which we consider, the
swimming speed of a spherical Janus particle is independent of size with $|U| =
0.25$ for uniform mobility~\cite{RG1}, so that the optimal Janus torus is 31\%
faster.

			For the Janus torus, the swimming speed in the case of nonuniform mobility is lower than for uniform mobility, and there is a much sharper peak around the optimum than for the other cases. This is because concentration gradients after $\theta = \pi$ cannot contribute to the swimming velocity. The product concentration (Fig.~\ref{fig:6}, left) shows that there are large surface concentration gradients on the inside of the torus, around $\theta = \pi$, which do not all contribute to $U$. The position and extent of these concentration gradients depend on $s_0$, giving rise to the sharper peak for the nonuniform mobility.
						
\section{\label{sec:4}Discussion}
	
In this paper we computed the  analytical solution for the autophoretic motion of an axisymmetric torus in {a neutral solute}, in the purely diffusive limit and following  
the continuum framework of {Golestanian et
al~\cite{golestanian2005}}. Using separation of variables in toroidal
coordinates, we solved for the product concentration outside of the torus
{in the case of axisymmetric surface activity}. In the central hole,
geometric confinement leads to locally higher product concentrations,
{creating surface concentration gradients that drive slip flow even
in the absence of chemical patterning.} Upon optimizing swimming and pumping
with respect to fixed surface area, we found that maximal confinement is not
optimal, demonstrating a qualitative difference to {the optima of
other} autophoretic systems {driven purely by geometric
confinement}~\cite{SM2,ML1}. 
		
For the {chemically uniform} torus (i.e.~$A, M = \pm 1$)  symmetry prevents
swimming, {but} concentration gradients from the central hole
{generate a pumping flow}. We characterized
this pumping flow by the strength of the resulting stresslet singularity. Fixing
the surface area, we varied the aspect ratio $s_0$; for vanishing central holes,
$s_0 \to 1$, the concentration is very large around $\theta = \pi$,
however these large local concentrations do not necessarily lead to large
concentration gradients {across the torus surface}. Meanwhile, for large
$s_0$, the distance between opposing ends of the torus means confinement
weakens, correspondingly there are smaller concentration variations over the
surface of the torus. These effects lead to an optimal torus with nonvanishing
central hole at $s_0 = 3.31$, which maximizes {the strength of the
pumping flow}, {and yields the stresslet with the highest} coefficient~$k$.
		
We next considered nonuniform tori where the up-down symmetry of the system is
broken, resulting in swimming. We examined a smoothed approximation to the
activity of a Janus torus, as such first-mode regularizations of Janus particles
are used in the study of spherical Janus particles~\cite{YI1}, and then we
compared it against a full Janus torus where we took both the surface activity
and mobility to have step discontinuities at $\theta = 0$ and $\pi$. In
maximising the swimming velocity of the nonuniform tori with respect to fixed
surface area, we found (Fig.~\ref{fig:7}) that (b) the optimal smoothed-activity
torus, has a larger aspect ratio than the optimum for the full Janus torus with
the same uniform mobility, (c).  The optimal swimmers, (b), (c), and (d), all
have smaller aspect ratios than (a) the optimal phoretic pump; unlike the pump,
swimmers must not only maximize surface concentration gradients, but also
minimize drag. The smaller aspect ratios result from the balance of these two
effects.

{While our study  focused on neutral solutes and 
self-diffusiophoresis,  if suitably modified the results may be
applied to other phoretic mechanisms. For instance, the auto-thermophoretic torus may
be studied provided that both the mechanism of heating is axisymmetric, and
additionally the interior problem of heat conduction through the torus is
solved. The slip-velocity is then the gradient of the temperature field at the
torus surface~\cite{kroy2016hot}, and the flow solution proceeds as before. For
electrophoresis, the situation is more complex as in general  it also involves
charge transport through the material; slip velocity would then be proportional to the electric field on the
torus surface~\cite{anderson1989colloid}, with activity corresponding to
electrical current injected into the flow at the surface~\cite{RG1}.}
	
The torus is arguably the simplest system that naturally exhibits geometric
confinement, an important property that can influence the motion of
diffusiophoretic particles leading to physical effects such as boundary
steering~\cite{das2015boundaries,simmchen2016topographical} and dynamic self
assembly~\cite{wykes2016dynamic}. Our results provide a means of probing
confinement effects analytically, demonstrating key differences between pumps
and swimmers.\begin{acknowledgements}
	
This work was undertaken as part of EPSRC-funded UROP project. EL is supported
in part by the European Union through a Marie Curie CIG grant and an ERC Consolidator grant. TDM-J is
supported by a Royal Commission for the Exhibition of 1851 Research Fellowship.
The authors are thankful to S\'{e}bastien Michelin for fruitful discussions,
{and to the anonymous referees for useful feedback.}
\end{acknowledgements}



\end{document}